**Reorientation of Sputnik Planitia implies a Subsurface Ocean on Pluto**

F. Nimmo*[1], D.P. Hamilton[2], W.B. McKinnon[3], P.M. Schenk[4], R.P. Binzel[5], C.J. Bierson[1], R.A. Beyer[6], J.M. Moore[6], S. A. Stern[7], H. A. Weaver[8], C. Olkin[7], L. A. Young[7], K. E. Smith[6] and the New Horizons Geology, Geophysics and Imaging Theme Team

[1]Department of Earth and Planetary Sciences, University of California Santa Cruz, Santa Cruz CA 95064; [2]Department of Astronomy, University of Maryland, College Park MD 20742; [3]Department of Earth and Planetary Sciences and McDonnell Center for the Space Sciences, Washington University in St Louis, St. Louis MO, 63130; [4]Lunar and Planetary Institute, Houston TX 77058; [5]Department of Earth, Atmospheric and Planetary Sciences, Massachusetts Institute of Technology, Cambridge MA 02139; [6]National Aeronautics and Space Administration (NASA) Ames Research Center, Moffett Field, CA, 94035; [7]Southwest Research Institute, Boulder CO, 80302; [8]Johns Hopkins University Applied Physics Laboratory, Laurel MD, 20723

**The deep nitrogen-covered Sputnik Planitia (SP; informal name) basin on Pluto is located very close to the longitude of Pluto's tidal axis[1] and may be an impact feature[2], by analogy with other large basins in the solar system[3,4]. Reorientation[5-7] due to tidal and rotational torques can explain SP's location, but requires it to be a positive gravity anomaly[7], despite its negative topography. Here we argue that if SP formed via impact and if Pluto possesses a subsurface ocean, a positive gravity anomaly would naturally result because of shell thinning and ocean uplift, followed**

**by later modest $N_2$ deposition. Without a subsurface ocean a positive gravity anomaly requires an implausibly thick $N_2$ layer (>40 km). A rigid, conductive ice shell is required to prolong such an ocean's lifetime to the present day[8] and maintain ocean uplift. Because $N_2$ deposition is latitude-dependent[9], nitrogen loading and reorientation may have exhibited complex feedbacks[7].**

The SP basin is 3.5 km below its surroundings (Figure 1) and is filled with a convecting layer of nitrogen ice, thought to be ~3-10 km thick[10,11]. This structure would yield a strongly negative gravity anomaly (Extended Data Fig 1); to generate a present-day positive gravity anomaly either a much thicker $N_2$ layer or some other source of extra mass at depth would be required.

Stereo topography[1,2] suggests a present-day elliptical shape of 1300 x 900 km. The topography resembles that of other large degraded impact basins such as Hellas[3] or Caloris[4] and includes a sharp rim (informally, Cousteau Rupes) to the north-east[1]. Elevated topography beyond the basin rim might represent ejecta, but a distinct ejecta blanket is not visible in images[1], perhaps because of modification by Pluto's ongoing surface geological activity. The centre of the SP ellipse is at about 175° E, 18°N, or about 400 km from the tidal axis. A randomly-placed point has only a 5% chance of being this close or closer to either tidal axis.

If SP formed during an impact then its initial depth $d_0$ was probably about 7 km [Methods], based on the depths of unrelaxed basins on Iapetus and the Moon[12], with uncertainties introduced by the low velocities of Pluto impactors[13]. The horizontal scale of SP suggests that a thickness of tens of kilometers of ice was removed during impact, and that impact-driven uplift of an ice-ocean interface (if present) probably occurred[14]. This uplift is important because it represents a large mass excess (Extended Data Figure 1). On the Moon a combination of impact-driven uplift of dense mantle material and later surface addition of lavas after the crust has cooled and strengthened results in impact basins showing a positive gravity anomaly[15-17]. We argue below that an analogous set of processes occurred at SP.

.

If Sputnik Planitia represents a positive gravity anomaly, tidal and rotational torques will have reoriented it towards the tidal axis. The calculated reorientation is mainly equatorwards (Figure 1c) and depends on the amplitude of the positive gravity anomaly, parameterized by the dimensionless parameter $Q$ [Methods]. Because of Pluto's slow spin rate, the stabilizing effect of any remnant rotational bulge is small and equatorwards reorientation can occur for modest (few tens of mGal) positive gravity anomalies. A 20º reorientation increases the probability of SP's initial location being that close to a tidal axis to 23%. Our calculations are conservative because they neglect the role of the ejecta blanket, silicates contained in the impactor, and decoupling of the shell from the silicates underneath, all of which will serve to increase reorientation [Methods]. Conversely, if SP represents a present-day negative gravity anomaly it must have formed closer to the equator [Methods].

We now calculate likely gravity anomalies at SP. If no ocean was present, uplift of the silicate interior is unlikely to have happened because of its rigidity and great depth[14] (assuming a differentiated body). In this case, we assume that deposition of $N_2$ of thickness $L$ took place at a later epoch by which time the crust had an elastic thickness $T_e$. Thermal evolution models predict that $T_e$ always exceeds 40 km, depending mainly on when SP formed[18]. Given $d_0$ and the present-day topography $h$, the load thickness $L$ and the resulting gravity anomaly $\Delta g$ can be calculated (Figure 2a; Methods). For basins with initial depths in the range 0-7 km, positive gravity anomalies only occur with $N_2$ loads > 40 km thick and $T_e$ values < 15 km (so that the space required by the $N_2$ can be accommodated). The required $N_2$ thickness is much larger than that inferred[10,11] and the $T_e$ value is smaller than predicted[18]. The large negative gravity anomaly generated by the present-day 3.5 km negative topography is hard to overcome with $N_2$ loading alone.

If a subsurface ocean is present, the post-impact, pre-loading state is assumed to be isostatic, resulting in a thinned shell beneath the basin[14,16]. The dense water beneath the basin thus provides an additional positive contribution to the overall gravity. For example, Figure 2b shows that with an ocean an $N_2$ layer 7 km thick can generate a +32 mGal gravity anomaly for $T_e$=70 km. These values are consistent with the available constraints.

If SP is a positive gravity anomaly at the present-day, Figure 2 suggests that a subsurface ocean with a thinned shell beneath the basin provides a viable explanation. Such a configuration will be smoothed out by lateral flow of the ice[19] at a rate dependent on the ice viscosity and the shell thickness $t_c$. Figure 3 shows that the configuration can be maintained for 4 Gyr as long as the base of the ice shell is cold, 180-250 K depending on shell thickness. Such low temperatures can be achieved with an ammonia- and/or methanol-bearing ocean[20] (ammonia is present in the Pluto system[21]) and imply a conductive shell, a large fraction of which will behave elastically. A conductive shell also transfers heat sufficiently slowly that a subsurface ocean can survive to the present day[8,22]. Preferential refreezing of the thinned portion of the shell could remove shell thickness contrasts. However, the thinned portion is capped by solid $N_2$, which has a much lower thermal conductivity than ice[23] and – even if convecting[10] – can provide sufficient insulation to prevent the thinned shell from refreezing [Methods].

Rather than uplift of liquid water underlying the ice shell, uplift of mantle material, dense, solid ice II, silicate-rich ice or reduced-porosity ice might instead be contributing to $\varDelta g$. We argued above that the first possibility was unlikely. We do not favour the second alternative because the presence of ice II implies strongly compressional tectonics[20,22], for which there is no evidence[1]. Theoretical models[24] predict that silicate-rich ice, if present, should be found at the surface, because of the low temperatures, while deeper ice should be silicate-free. This is opposite to the required distribution. An impact-induced porosity reduction of 10% would need to extend to a depth of 70 km to compensate the basin, but for SP-size basins the porosity effect on gravity is likely overwhelmed by uplift of the underlying material[14,25]. Although impact-driven ocean uplift is expected for an SP-forming impact[14], further work will be required to definitively exclude these other alternatives.

An alternative hypothesis[26] suggests that the SP basin formed by early loading of $N_2$ ice and reorientation as Pluto's spin state evolved to synchronous. In this hypothesis $N_2$ was subsequently removed from SP; this removal would cause >10° of polewards motion [Methods] and affect $N_2$ deposition. This prediction of polewards motion is opposite to that shown in Figure 1; since reorientation[27] and load removal cause tectonic stresses, mapping of tectonic features[7] should be able to test which of these hypotheses is correct.

If Pluto contains a cold (likely $NH_3$-bearing) liquid ocean, several further issues arise. The predicted slow re-freezing of a Plutonian ocean results in isotropic extensional stresses[8,22], in agreement with the tectonic features observed[1] . The requirement for shell thinning to have occurred allows numerical models to probe the present-day shell thickness[14]. A rigid, conductive shell could be reconciled with putative cryovolcanic surface features[1] by appealing to ocean pressurization caused by progressive thickening of the ice shell[28]. Various Kuiper Belt Objects of somewhat similar sizes and densities (bulk compositions) to Pluto are known[29]; among these bodies, subsurface oceans are likely a common phenomenon.

**Acknowledgements** New Horizons was built and operated by the Johns Hopkins Applied Physics Laboratory (APL) in Laurel, Maryland, USA, for NASA. We thank the many engineers, flight controllers and others who have contributed to the success of the New Horizons mission and NASA's Deep Space Network (DSN) for a decade of excellent support to New Horizons. We thank Brandon Johnson for helpful discussions on impact physics, Jack Conrad for cryovolcanism calculations, and two anonymous reviewers for their comments.

**Author Contributions** DPH originated the reorientation hypothesis, FN developed the subsurface ocean scenario and carried out the bulk of the calculations; CJB calculated the effect of realistic basin geometries and ejecta blanket. PMS and RAB provided the stereo topography. All authors read or commented on the MS.

**Author Information**

Reprints and permissions information is available at www.nature.com/reprints.

The authors declare no competing financial interests.

Correspondence and requests for materials should be addressed to FN.

GGI Team members: J.M. Moore[1], W.B. McKinnon[2], J.R. Spencer[3], R. Beyer[1], R.P. Binzel[25], M. Buie[3], B. Buratti[4], A. Cheng[5], D. Cruikshank[1], C.Dalle Ore[1], A. Earle[25], R. Gladstone[6], W. Grundy[7], A.D. Howard[8], T.Lauer[9], I. Linscott[10], F. Nimmo[11], C. Olkin[3], J. Parker[3], S. Porter[3], H. Reitsema[12], D. Reuter[13], J.H. Roberts[5], S. Robbins[3], P.M. Schenk[14], M. Showalter[15], K. Singer[3], D. Strobel[16], M. Summers[17], L. Tyler[10], H. Weaver[5], O.L. White[1], O.M. Umurhan[1], M. Banks[18], O. Barnouin[5], V. Bray[19], B. Carcich[20], A. Chaikin[21], C. Chavez[1], C. Conrad[3], D. Hamilton[22], C. Howett[3], J. Hofgartner[20], J. Kammer[3], C. Lisse[5], A. Marcotte[5], A. Parker[3], K. Retherford[6], M. Saina[5], K. Runyon[4], E. Schindhelm[3], J. Stansberry[23], A. Steffl[3], T. Stryk[24],. H. Throop[3], C. Tsang[3], A. Verbiscer[8], H. Winters[5], A. Zangari[3], S.A. Stern[3], H.A. Weaver[5], C.B. Olkin[3], L.A. Young[3], K.E. Smith[1]

[1]National Aeronautics and Space Administration (NASA) Ames Research Center, Moffett Field, California 94035, USA. [2]Department of Earth and Planetary Sciences and McDonnell Center for the Space Sciences, Washington University in St Louis, Saint Louis, Missouri 63130, USA. [3]Southwest Research Institute, Boulder, Colorado 80302, USA. [4]NASA Jet Propulsion Laboratory, Pasadena, California 91019, USA. [5]Johns Hopkins University Applied Physics Laboratory, Laurel, Maryland 20723, USA. [6]Southwest Research Institute, San Antonio, Texas 78238, USA. [7]Lowell Observatory, Flagstaff, Arizona 86001, USA. [8]University of Virginia, Charlottesville, Virginia 22904, USA. [9]National Optical Astronomy Observatory, Tucson, Arizona 85719, USA. [10]Stanford University, Stanford, California 94305, USA. [11]Department of Earth and Planetary Sciences, University of California Santa Cruz, Santa Cruz, California 95064, USA. [12]B612 Foundation, Mill Valley, California 94941, USA. [13]NASA Goddard Space Flight Center, Greenbelt, Maryland 20771, USA. [14]Lunar and Planetary Institute, Houston, Texas 77058, USA. [15]The SETI Institute, Mountain View, California 94043, USA. [16]The Johns Hopkins University, Baltimore, Maryland 21218, USA. [17]George Mason University, Fairfax, Virginia 22030, USA. [18]Planetary

Science Institute, Tucson, Arizona 85719, USA. [19]University of Arizona, Tucson, Arizona 85721, USA. [20]Cornell University, Ithaca, New York 14853, USA. [21]Arlington, Vermont 05250, USA. [22]University of Maryland, College Park, Maryland 20742, USA. [23]Space Telescope Science Institute, Baltimore, Maryland 21218, USA. [24]Roane State Community College, Oak Ridge, Tennessee 37830, USA. [25]Massachusetts Institute of Technology, Cambridge, Massachusetts 02139, USA.

**Figure Captions**

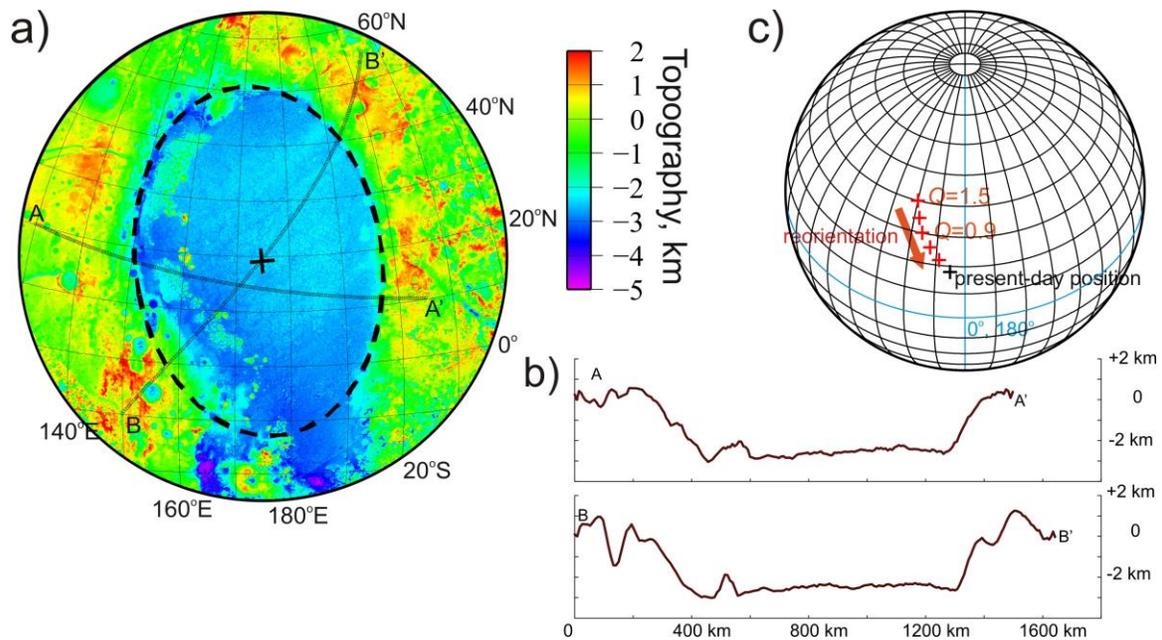

**Figure 1. Sputnik Planitia (SP) topography and reorientation a)** Stereo-derived topography of SP (using method described in ref. 1) with an ellipse with axes 1300x900 km superimposed. The ellipse centre and projection centre (Lambert equal area) are both 175ºE, 18ºN. **b)** Topographic profiles, locations shown in a). Point spacing was 8 km with 5-point averaging to reduce noise. **c)** Location of SP prior to reorientation (red crosses) as a function of dimensionless gravity anomaly *Q* (in increments of 0.3). A *Q* of 1.4 represents a nominal peak gravity anomaly Δ*g* of +31 mGal [Methods] and yields about 20º true polar wander. Orthographic projection centred at 180ºE, 45ºN.

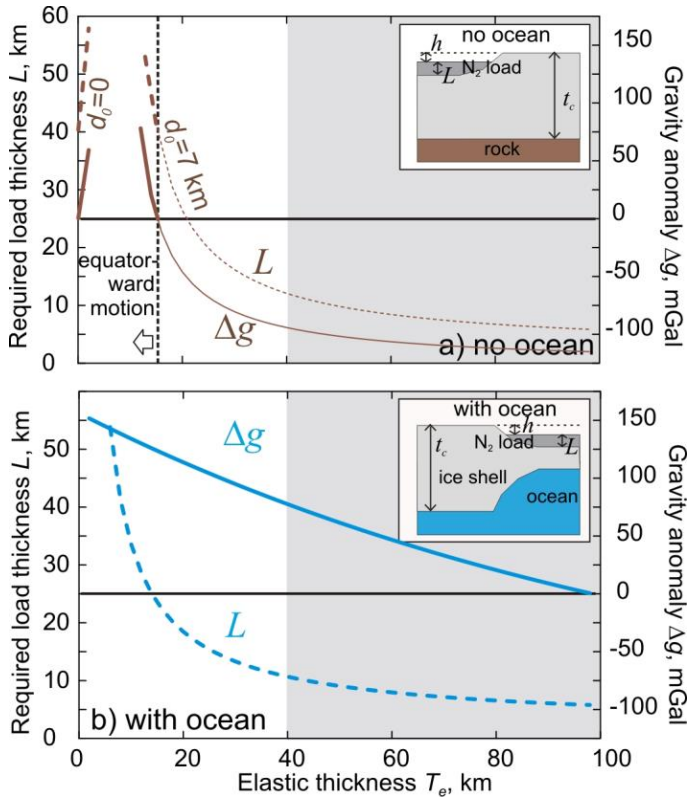

**Figure 2. Load thicknesses $L$ and resulting gravity anomalies $\Delta g$ for present-day Sputnik Planitia topography. a)** Case with no ocean. Equatorwards reorientation takes place if $\Delta g$ is positive. Initial basin depth is $d_0$; to generate the present-day negative topography ($h$=3.5 km) the deflection due to a load thickness $L$ is calculated using a thin-spherical-shell approach[30] (see Methods). Shaded region denotes estimated elastic thickness range[18]. The characteristic wavenumber of SP is taken to be $(4/3)\pi/D$ where $D$ is the diameter (=1000 km). Inset shows model geometry assumed. **b)** Case with ocean in which the pre-loading basin is isostatically compensated. Here $d_0$=7 km. The shell thickness $t_c$ is taken to be $2T_e$ for calculating the gravity contribution of the water; this can be justified *a posteriori* by the requirement for a cold, conductive shell (Fig 3).

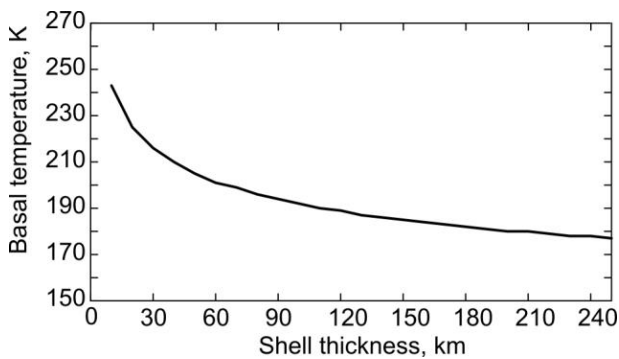

**Figure 3 | Basal shell temperature required to maintain a thinned shell for 4 Gyr.**
Timescale calculated using ref. 19 assuming a Newtonian viscosity of $10^{14}$ Pa s at 270 K
and an activation energy of 50 kJ/mol [Methods]. A conductive temperature profile was
assumed with the surface at 40 K.

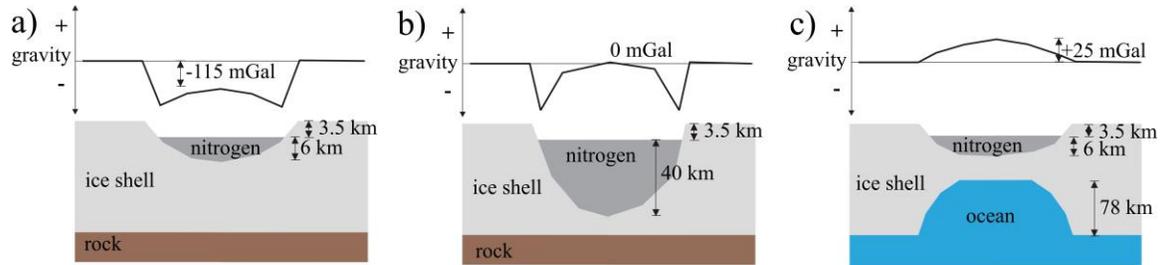

**Extended Data Figure 1.** Schematic models of SP structure and associated gravity
profiles. The peak gravity anomaly is calculated using the flat-plate formula $2\pi G\Delta\rho h$ for
each layer, where $h$ represents the thickness, $\Delta\rho$ the lateral density contrast and the
densities of $H_2O$ ice, water and $N_2$ ice are 0.92, 1.0 and 1.0 g/cc (ref 23), respectively. In
panel c) the gravitational contribution of the ocean is reduced due to upwards attenuation
with a shell thickness of 150 km [see Methods]. Either a >40 km thick nitrogen layer or
an uplifted ocean could result in a present-day positive gravity anomaly at SP. Panel c) is
similar to the inferred structure of lunar mascon basins, which also show positive gravity
anomalies (refs 15,16).

## Methods

*Reorientation.* To calculate the reorientation due to SP loading we follow the methods of
ref. 27 with one exception. For a tidally-distorted, slowly-rotating synchronous satellite,
the ratio of the non-normalized hydrostatic degree-two gravity coefficients $J_2/C_{22}$=10/3.
However, since Pluto is the primary, it experiences less tidal distortion and the coefficient
ratio is correspondingly higher, $\gg$ 14.3 (ref. 7). As a result, we generalize equation (39)
of ref. 27 as follows:

$$Q\sin 2q_L\cos\left(f_R - f_L\right) = \sin 2q_R\left[1 + \frac{f\cos^2 q_T}{\sin^2 q_R}\right] \quad (1)$$

Here $Q$ is the dimensionless load size, $\theta$ and $\phi$ are colatitude and longitude, respectively,
and the subscripts $L,T$ and $R$ refer to the final location of the load and the initial locations
of the tidal axis and the rotational axis in the final reference frame. Here $f$ is defined as
$f$=3$m/(M+m)$, where $m$ and $M$ are the masses of the tide-raising body (Charon) and Pluto,
respectively, such that for a synchronous satellite orbiting a massive planet, $f$=3 (yielding
equation 39 of ref. 27) while for a purely rotationally-distorted body $f$=0. With this
modification the reorientation due to an imposed load $Q$ may be calculated. For
simplicity, we assume that reorientation occurs as a single event, though in reality it may
have consisted of progressive motion.

For $Q$=1 and $f$=0.327 (appropriate to Pluto) we find that $\theta_T$=102.5°, $\phi_{T=}$193.1°, $\theta_R$=13.6° (this is the amount of true polar wander,TPW) and $\phi_R$=169.6°. A TPW of 20° requires $Q$=1.4. The initial position of the load in the final reference frame may then be calculated using spherical triangles or by diagonalizing the moment of inertia tensor (ref. 27); for $Q$=1 the load is initially located at 31.6° latitude, 163.2° longitude in the final reference frame.

A basin of constant depth $h$ and angular radius $\psi$ yields the following dimensionless load $Q$ (ref. 6):

$$Q = \frac{3\pi G h \rho \cos\psi \sin^2\psi}{R\Omega^2 \Delta k_2} = \frac{3}{2}\frac{p\Delta g \cos\psi \sin^2\psi}{R\Omega^2 \Delta k_2} \qquad (2)$$

where $G$ is the gravitational constant, $\rho$ is the density of the material, $R$ and $\Omega$ are the radius and rotation angular frequency of Pluto and $\Delta k_2$ is the difference between the fluid Love number and the actual Love number (this quantity describes the size of the remnant bulge, which opposes reorientation). The numerator depends on the size of the load and the denominator represents the remnant bulge size. The size of the remnant bulge depends on $\Delta k_2$ and the rotation rate at which the bulge was "frozen in". Existing shape observations show no evidence of a remnant bulge[31] and the establishment of Pluto's present-day spin rate probably took a few Myr[31], whereas cooling of the interior and freezing in of a remnant bulge probably took tens to hundreds of Myr[8,22]. We therefore take the relevant rotation rate to be that of the present day. The second equality introduces the peak gravity anomaly $\Delta g$ associated with the basin. For a parabolic basin (as we assume for SP), the peak gravity is the same as for the constant-depth case, but the corresponding value of $Q$ is reduced by a factor $p \approx 0.5$ because the mean basin depth is smaller. We take $R$=1188 km, $p$=0.5, $\Omega$=1.14x10$^{-5}$ rad s$^{-1}$, and $\psi$=24° ($D$=1000 km). For a Pluto with a 50 km thick elastic lithosphere $\Delta k_2$=0.16 (see below) in which case equation (2) yields $\Delta g$=22 $Q$ mGal. A larger $\Delta k_2$ (larger remnant bulge) would require a larger gravity anomaly to get the same amount of reorientation.

Our calculated degree of reorientation is likely conservatively small, for three reasons. First, if present, an ejecta blanket will reduce the size of the original negative gravity anomaly associated with the basin (yielding $p \approx 0.3$). Second, the basin-forming impactor probably contained some silicates, so any impactor material incorporated into the ice shell will provide a positive contribution to gravity. Third, a decoupled ice shell is likely to reorient more than a solid body. However, for our argument the degree of reorientation is less important than the sign: only a basin exhibiting a positive gravity anomaly will experience equatorwards reorientation.

*Polewards Motion.* For a load near the tidal axis and for a body (like Pluto) which is primarily rotationally distorted, we can approximate equation (1) as $Q\sin 2\theta_L \approx \sin 2\theta_R$ with $\theta_L$=72° for present-day SP. The present-day gravity anomaly in the absence of a subsurface ocean is about -115 mGal (Extended Data Figure 1). Using the present-day rotation period and setting $\Delta k_2$=1 to represent the largest likely remnant bulge (the real value is probably considerably smaller; see below) and with $\Delta g$=-115 mGal , equation (2) shows that the corresponding value of $Q$ is -0.8. This in turn implies a polewards reorientation $\theta_R$ of about 14°, and an original (pre-reorientation) latitude of 4°. A smaller remnant bulge would result in more reorientation. If SP is a negative gravity anomaly at the present day, or if mass was removed after its equilibrium position was established, SP should have experienced large polewards reorientation, because the stabilizing effect of the rotational remnant bulge is small.

*Loading Calculations*. Consider first a basin that is initially isostatically-compensated by an uplifted root (the with-ocean case), so that the initial gravity anomaly is ~0. The initial uplift $r$ is given by $r=d_0 \, \rho_c/(\rho_m-\rho_c)$ where $\rho_m$ and $\rho_c$ are the density of water and ice, respectively, and $d_0$ is the depth of the basin after rebound. Assuming that an initially unstressed elastic layer develops after the rebound is complete, subsequent loading results in deflection. Taking the load thickness to be $L$, the deflection $w$ (positive downwards) and the final basin negative topography $h$, we have

$$h=d_0+w-L \qquad (3)$$

For a load described by a single spherical harmonic degree $n$, the required load thickness $L$ for a given $h$ can then be obtained via

$$L = \frac{(h-d_0)(C'_n+1)}{\left(\frac{\rho_L-\rho_c}{\rho_c}C'_n-1\right)} \qquad (4)$$

Here $\rho_L$ is the load density, $C'_n = \frac{r_c}{r_m-r_c}C_n$ where $C_n$ is the degree of compensation[30] which depends on the elastic thickness and we have modified the definition from ref. 30 to avoid singularities arising when $\rho_m=\rho_c$. In the rigid limit there is no deflection, $C'_n=0$ and equation (4) yields the correct answer [$L=d_0-h$]. In the isostatic limit $C'_n = \frac{\rho_c}{\rho_m-\rho_c}$ and again the correct answer is recovered [$L=(d_0-h)\rho_m/(\rho_m-\rho_L)$], yielding a much larger load thickness.

The post-loading peak gravity anomaly is given by

$$\mathrm{D}g = 2\rho G\Big(-[h+L]\,r_c + L\,r_L + [r-w]\big[\,r_w - r_c\,\big]e^{-kt_c}\Big) \qquad (5)$$

The final term in equation (5) represents the positive gravity contribution of the uplifted dense water. Here the factor exp(-$kt_c$) is due to upwards attenuation of the gravity anomaly owing to the finite shell thickness $t_c$. We take $t_c=2\,T_e$.

Next we consider a basin overlying a flat ice-silicate interface (no-ocean case). The depth after any initial (pre-loading) flexure is taken to be $d_0$. The required load thickness can again be obtained from equation (4) where in this case $C'_n$ is calculated by setting $\rho_m=\rho_c$ (because there is no contribution from a higher-density layer at depth). Again, the correct answer is recovered in the rigid and isostatic limiting cases. In this case the peak gravity anomaly is then simply

$$\mathrm{D}g = 2\rho G\Big(-\big[h+L\big]\,r_c + L\,r_L\Big) \qquad (6)$$

We calculate $C_n$ using eq. 27 of ref. 30. We convert from wavenumber $k$ to spherical harmonic degree $n$ by using $n \gg kR$. The Young's modulus of ice is 9 GPa, densities of water ice, water and $N_2$ ice are taken to be 0.92, 1.0 and 1.0 (ref. 23) g/cc, respectively. Incorporation of $NH_3$ into the ice could in theory reduce its effective rigidity, but during slow freezing $NH_3$ will be excluded from the crystallizing ice[32].

In reality, SP loading consists of contributions from multiple wavenumbers. To determine the dominant wavenumber, we calculated the flexural deflection of a parabolic basin using the approach of ref. 33 and determined that the maximum deflection is well-approximated by an effective wavenumber $k=4\pi/3D$, where $D$ is the basin diameter.

*Lateral flow of the shell.* The timescale for lateral flow of the shell is calculated using the approach of ref. 22 which gives the relaxation timescale $\tau$:

$$t = \frac{\eta_b}{g \Delta \rho \delta^3 k^2}$$

where $\eta_b$ is the basal viscosity, $k$ is the wavenumber as before, $\delta$ is the effective layer thickness in which flow occurs and $\Delta\rho$ is the ice-water density contrast. The basal viscosity depends on the reference viscosity and the activation energy $Q_a$, and for a shell in which conductivity varies as $1/T$, $\delta$ is given by

$$\delta = \frac{R_g T_b t_c}{Q_a \ln(T_b / T_s)}$$

where $R_g$ is the gas constant and $T_b$ and $T_s$ are the basal and surface temperatures.

*Size of remnant bulge.* The size of the remnant bulge[27,34] is assumed to depend on the quantity $k_{2f}$-$k_2$, where $k_{2f}$ is the Love number after all stresses have relaxed and $k_2$ is the present-day Love number. A body which is fluid at the present day has no remnant bulge ($k_{2f}$-$k_2$=0) while a body which is infinitely rigid now ($k_2$=0) has the largest possible remnant bulge, the size of which depends on the density structure and initial rotation rate. We use the method of ref. 35 to calculate the Love numbers and assume that the body is spherically symmetric. We assume that Pluto's silicate interior has remained rigid and unrelaxed at all timescales and has an outer radius of 842 km, a rigidity of 100 GPa and a density of 3.5 g/cc. The overlying $H_2O$ layer has a mean density of 0.95 g/cc and an outer radius of 1188 km. In the presence of an elastic ice shell 50 km thick with a shear modulus of 3 GPa, $k_2$=0.28, while in the absence of such a shell $k_{2f}$=0.44. The fact that $k_{2f}$-$k_2 \gg k_2$ implies that the remnant bulge and present-day bulge are of comparable magnitude. Our assumption of a rigid silicate core is based on thermal evolution calculations[8]; if the core were instead strengthless at all timescales, the Love numbers increase to $k_2$=0.52 and $k_{2f}$=0.75, respectively.

*Initial depth of SP basin.* Pluto's radius is close to the geometric mean of the radii of Iapetus ($R$=734 km) and the Moon ($R$=1738 km). Thousand-km diameter, apparently unrelaxed basins exist on the latter two bodies[12] with Iapetus basins approaching 10 km in depth and lunar basins about a factor of two shallower. A similar-scale unrelaxed basin on Pluto might therefore be expected to be $\gg$ 7 km deep. The corresponding isostatic ocean uplift would be 80 km. Expected impact velocities on Pluto are lower even than on Iapetus, but the implications of these lower velocities for the initial depth:diameter ratio of the resulting basin are unclear[13].

The extent to which crust (shell) thinning and mantle (ocean) uplift occur in response to an impact depend on the diameter of the basin relative to the depth to the mantle/ocean[14,16]. On the Moon, with a mean crustal thickness of about 35 km, mantle uplift occurs for basins with diameters in excess of 220 km (refs. 25,36). Assuming that this same ratio applies to Pluto, a 1000 km diameter basin would be expected to generate ocean uplift for shells thinner than about 160 km. This expectation is confirmed by numerical models[14] which show that uplift occurs for ice shell thicknesses less than ~180 km. A chondritic Pluto might have a present-day shell thickness similar to this value[8,24], while in the past the shell will have been thinner and uplift correspondingly more likely to have occurred.

*Insulating effect of $N_2$*. Consider a reference shell of thickness $t_c$ and effective thermal conductivity $k_c$. It may be compared with a thinned shell of total thickness $t_c'$ containing a layer of lower conductivity ice $k'$ of thickness $L$. For the heat fluxes across the two shells to be equal, the required thickness of the insulating ice $L$ can be shown to be

$$L = \frac{k'}{k_c - k'}\left(t_c - t_c'\right)$$

We note that this analysis neglects any melting at the base of the $N_2$ layer. Water ice exhibits a temperature-dependent thermal conductivity given by $651/T$ (ref. 37). The effective thermal conductivity $k_c$ over the temperature range 40-240 K is then 5.8 $Wm^{-1}K^{-1}$. By contrast, nitrogen ice at 50K has a thermal conductivity of 0.2 $Wm^{-1}K^{-1}$ (ref. 23). The effective thermal conductivity of the nitrogen will be increased if it is convecting. Based on the results of ref. 10, the Nusselt number of the convecting nitrogen is about 3, so that the effective nitrogen thermal conductivity $k' \approx 0.6$ $Wm^{-1}K^{-1}$.

For an initial basin depth of 7 km, the shell thinning after loading ($t_c$-$t_c'$) at SP will be about 70 km depending on the exact densities assumed and the amount of deformation. Thus, a nitrogen layer 8 km thick is sufficient to offset the increased heat flux due to the thinned shell. As a result, shell thickness variations can be maintained over geological timescales as long as an insulating $N_2$ ice layer persists. As shown in Fig 2, a layer this thick will yield a positive gravity anomaly of about +30 mGal, sufficient to cause reorientation.

*Code Availability*. Codes for the reorientation, loading and lateral flow calculations are available upon request from FN.